\documentclass{aa}  

\usepackage{graphicx}
\usepackage{natbib}
\bibpunct{(}{)}{;}{a}{}{,} 
\usepackage{txfonts}
\usepackage[
  breaklinks = true,
  colorlinks = true,
  urlcolor   = blue,
  citecolor  = blue,
  linkcolor  = blue,
]{hyperref}
\usepackage{xcolor}

\newcommand{\HB}{\textnormal{H}\beta}
\newcommand{\Ha}{\textnormal{H}\alpha}
\newcommand{\He}{\textnormal{H}\varepsilon}
\newcommand{\BLOS}{B_{\textnormal{LOS}}}
\newcommand{\CaII}{\textnormal{Ca}~\textsc{ii}}
\newcommand{\kms}{\textnormal{km s}^{-1}}
\newcommand*\mean[1]{\overline{#1}}

\begin{document}

    \title{Quiet Sun Ellerman bombs as a possible proxy for \\
    reconnection-driven spicules}
    \titlerunning{QSEBs as a possible proxy for reconnection-driven spicules}

    \authorrunning{M.O. Sand et al.}
    
    \author{Mats Ola Sand
          \inst{1, 2},
          Luc H. M. Rouppe van der Voort
          \inst{1, 2},
          Jayant Joshi
          \inst{3},
          Souvik Bose
          \inst{4,5,1,2},
          Daniel Nóbrega-Siverio
          \inst{6,7,1,2},
          Ana Bel\'en Gri\~n\'on-Mar\'in
          \inst{8,9,1,2}
          }

    \institute{
            Institute of Theoretical Astrophysics, University of Oslo, P.O. Box 1029 Blindern, N-0315 Oslo, Norway
            \email{m.o.sand@astro.uio.no}
        \and
            Rosseland Centre for Solar Physics, University of Oslo, P.O. Box 1029 Blindern, N-0315 Oslo, Norway
        \and
            Indian Institute of Astrophysics, II Block, Koramangala, Bengaluru 560 034, India
        \and
            Lockheed Martin Solar \& Astrophysics Laboratory, Palo Alto, CA 94304, USA
        \and
            SETI Institute, 339 Bernardo Ave, Mountain View, CA 94043, USA
        \and
            Instituto de Astrofísica de Canarias, 
            38205 La Laguna, Tenerife, Spain
        \and
            Universidad de La Laguna, Dept. Astrofísica, 
            38206 La Laguna, Tenerife, Spain
        \and
            Institute for Solar Physics, Department of Astronomy, Stockholm University, Albanova University Centre, SE-106 91 Stockholm, Sweden
         \and 
            Swedish National Space Agency, SE-171 04, Solna, Sweden
            }

    \date{Received 16 November 2024; accepted 6 April 2025}
 
    \abstract
    {
    Spicules are elongated, jet-like structures that populate the solar chromosphere and are rooted in the lower solar atmosphere.
    In recent years, high-resolution observations and advanced numerical simulations have provided insights into their properties, structures, and dynamics. 
    However, the formation mechanism of spicules, particularly the more dynamic type II spicules, which are primarily found in the quiet Sun and coronal holes, remains elusive.
    }
    {
    This study explores whether quiet Sun Ellerman bombs (QSEBs), which are ubiquitous small-scale magnetic reconnection events in the lower atmosphere, are linked to the formation of type II spicules.
    }
    {
    We analysed a high-quality 40-minute time sequence acquired with the Swedish 1-m Solar Telescope.
    \(\HB\) data were used to observe QSEBs and spicules, while spectropolarimetric measurements in the photospheric \ion{Fe}{I}~6173~\AA\ line provided line-of-sight magnetic field information.
    We employed \(k\)-means clustering to automatically detect QSEBs and explored their potential association with spicules.
    }
    {
    We identified 80 clear cases where spicules occurred soon after the QSEB onset and not later than 30 s after the ending of the QSEBs. 
    In all these instances, the events involved type II spicules, rapidly fading from the images. 
    The footpoints of the spicules seemed to be rooted in QSEBs, where the onset of QSEBs often preceded the formation of the associated spicules. 
    In addition to these clear cases, we found around 500 other events that hinted at a connection but with some ambiguities.
    The combined clear and ambiguous cases constitute 34\% of the total detected QSEBs and a smaller percentage of the spicules in our dataset.
    }
    {
    Our findings suggest that a fraction of the type II spicules originate from QSEBs, supporting magnetic reconnection as a potential driving mechanism.
    In this context, QSEBs and spicules represent the conversion of magnetic energy into thermal and kinetic energy, respectively.
    We suggest that an observational programme including multiple Balmer lines would likely detect more unambiguous connections between QSEBs and spicules.
    }


    \maketitle


\section{Introduction}
\label{introduction}

    There has long been a debate about whether magnetic reconnection may drive spicules.
    Spicules are thin, highly dynamic jets of chromospheric plasma that shoot out from the lower atmosphere of the Sun, and the hypothesis of spicules forming from magnetic reconnection was first proposed by \citet{1969PASJ...21..128U} for the ``classical'' spicules \citep{1968SoPh....3..367B}.
    Today, based on observed dynamic behaviour, spicules are generally separated into type~I and type~II spicules.
    Type~I spicules follow a parabolic trajectory, while type~II spicules only rise before they swiftly fade out of the images when observed in wideband chromospheric images \citep[due to heating;][]{2007PASJ...59S.655D}.
    Type~I spicules have shown to be driven by magnetoacoustic shocks \citep{2006ApJ...647L..73H, 2007ApJ...655..624D}, while the much more dynamic type~II spicules became the candidate for the reconnection hypothesis.
    However, despite the ubiquity of type~II spicules \citep[dominating the quiet Sun and coronal holes;][]{2012ApJ...759...18P}, earlier research shows no conclusive observational evidence of magnetic reconnection driving spicules, and we lack consensus on their driving mechanism.
    Regarding their driving mechanism, radiative-MHD simulations suggest that type~II spicules can be driven by the release of amplified magnetic tension \citep{2017ApJ...847...36M,2017Sci...356.1269M,2020ApJ...889...95M}.
    Observational studies also argue that these spicules may form from magnetic reconnection between emerging and preexisting magnetic fields \citep[see, e.g.,][]{2019Sci...366..890S}, in a way similar to surges \citep{1996PASJ...48..353Y,2008ApJ...683L..83N,2016ApJ...822...18N}.
    Despite these arguments, there appears to be a general consensus on the fact that spicules are mass flows rooted in the lower atmosphere \citep[for a review, see][]{2012SSRv..169..181T}.
    Spicules appear to originate from the lower atmosphere, so searching for signs of energy release related to spicule formation in this region is natural.
    One form of impulsive and compact energy release in the lower atmosphere of active regions is known as Ellerman bombs \citep[EBs,][]{1917ApJ....46..298E}, and the literature agrees that these are events of strong-field magnetic reconnection in the photosphere \citep{2002ApJ...575..506G, 2004ApJ...614.1099P, 2007A&A...473..279P, 2006ApJ...643.1325F, 2008ApJ...684..736W}.
    EBs are characterised in the \(\Ha\) and \(\HB\) spectral lines by enhanced inner wing emission \cite[often referred to as moustache-like profiles;][]{1964ARA&A...2..363S} with an unaffected line core, and by their flame-like morphology and rapid variability 
    \citep{2011ApJ...736...71W, 
    2013JPhCS.440a2007R}.
    High-resolution observations in \(\Ha\) revealed that similar features also appear in the vicinity of bipolar magnetic regions in the quiet Sun \citep[quiet Sun Ellerman bomb-like brightenings (QSEBs);][]{2016A&A...592A.100R}.
    More recent studies, using the higher spatial resolution of \(\HB\), found that QSEBs can be seen everywhere in the quiet Sun, and they estimated that about half a million QSEBs could be present on the Sun at any time \citep{2020A&A...641L...5J, 2022A&A...664A..72J}.
    \citet{2024A&A...683A.190R} increased this estimate to 750\,000 after exploiting even higher spatial resolution in the \(\He\) spectral line.

    \begin{figure*}
    \sidecaption
      \includegraphics[width=12cm]{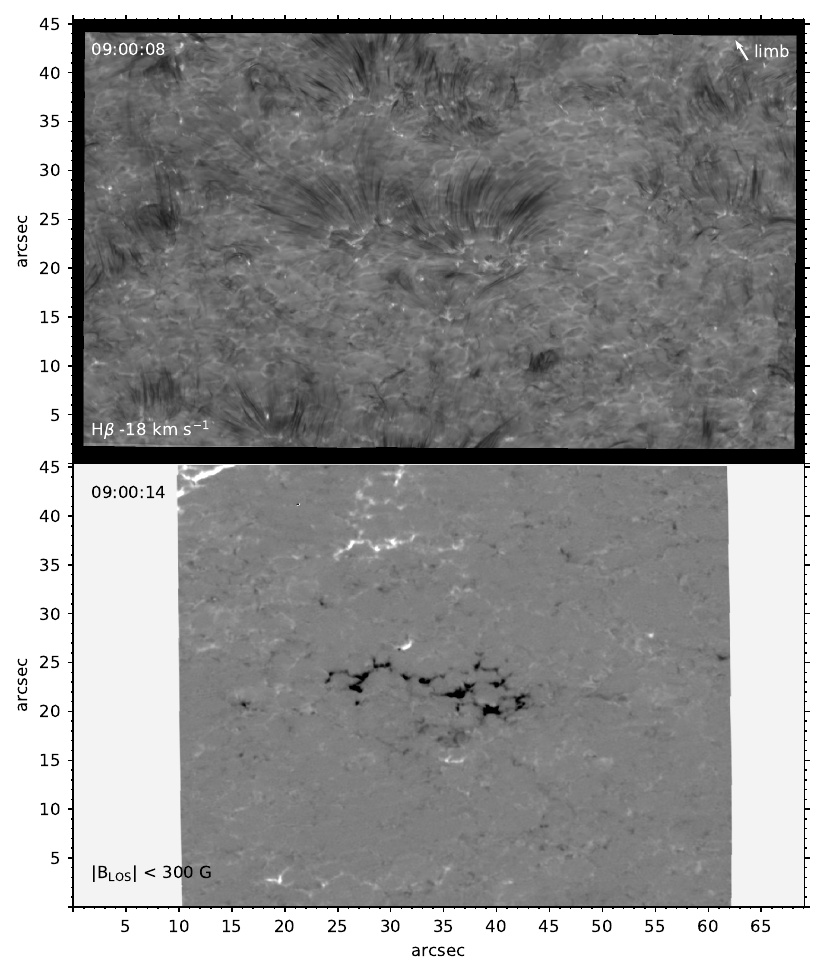}
        \caption{
        Overview of the quiet Sun region observed with SST on 25 July 2021 around 09:00 UT.
        Top: Blue wing of \(\HB\) at -18 \(\kms\) obtained with the CHROMIS instrument showing the spicule forest.
        The arrow in the top right corner shows the direction of the closest limb.
        Bottom: Line-of-sight photospheric magnetic field from the Milne-Eddington inversion of the \ion{Fe}{I}~6173~\AA\, line observed with CRISP.
        The map is saturated at $\lvert B_{\rm{LOS}} \rvert$~<~300~G.
        }
        \label{fig: data}
    \end{figure*}

    These new findings of the ubiquity of QSEBs hint at a potential relationship with type~II spicules, and suggest a way to establish the role of magnetic reconnection as a driver of the latter.
    We aim to see if we can determine how traceable reconnection events in the photosphere relate to the formation of type~II 
    spicules.
    To answer this, we combine the latest high-resolution observations and machine learning techniques to perform an automated search for QSEBs' potential connection to spicules.
    %


\section{Observations}
\label{observations}

    We used the high-resolution capabilities of the Swedish 1-m Solar Telescope \citep[SST;][]{2003SPIE.4853..341S} to observe a quiet Sun region on 25 July 2021 over 40 minutes, starting from 08:43 UT.
    The pointing was at solar \((X, Y) = (738\arcsec, 250\arcsec)\) with viewing angle \(\mu = 0.57\), where \(\mu = \cos{\theta}\) and \(\theta\) is the angle between the observer and the surface normal.
    The advantage of pointing more limbwards is that QSEBs are easier to detect due to their upright, flame-like morphology \citep{2013JPhCS.440a2007R}.
    The field of view (FOV), as shown in Fig.~\ref{fig: data}, contains several magnetic network patches with dense bushes of spicules originating from them.
    To track QSEBs and spicules, we used a single line \(\HB\) programme with the CHROMospheric Imaging Spectrometer instrument \citep[CHROMIS;][]{2017psio.confE..85S} with a cadence of \(7.2\)~s.
    The CHROMIS FOV covers \(66 \arcsec \times \, 42 \arcsec\) with a pixel scale of \(0 \farcs 038\), effectively capturing many of the small QSEB features in the photosphere.
    To cover the \(\HB\) signature for QSEBs, we sampled the line profile at 27 wavelength positions, between \(\pm 2.1\)~\AA\ from the nominal line centre; the step size was \(0.1\)~\AA\ between \(\pm 1\)~\AA\ and less refined in the outer wings to avoid blends.
    We also acquired \ion{Fe}{I}~6173~\AA\ spectropolarimetric observations with the CRisp Imaging SpectroPolarimeter instrument \citep[CRISP,][]{2008ApJ...689L..69S}. 
    We sampled the line at 13 line positions between \(\pm 0.32\)~\AA, with step size \(0.04\)~\AA, and at one continuum point at \(+ 0.68\)~\AA.
    We also sampled four line positions in \(\CaII\)~8542~\AA\ (not used in this study), giving a temporal cadence of \(18.6\)~s for the 2-line CRISP programme.
    We calculated the line-of-sight magnetic field (\(\BLOS\)) using the Milne-Eddington inversion code developed by \citet{2019A&A...631A.153D}.
    We measured the noise in the \(\BLOS\) frames by calculating the standard deviation in a small (\(50 \times 50\) pixels) and quiet region.
    The final value for the noise level, \(\sigma = 5.7\)~G, was then calculated as the average within this region throughout the time series.
        \begin{figure*}
        \centering
          \includegraphics[width=17cm]{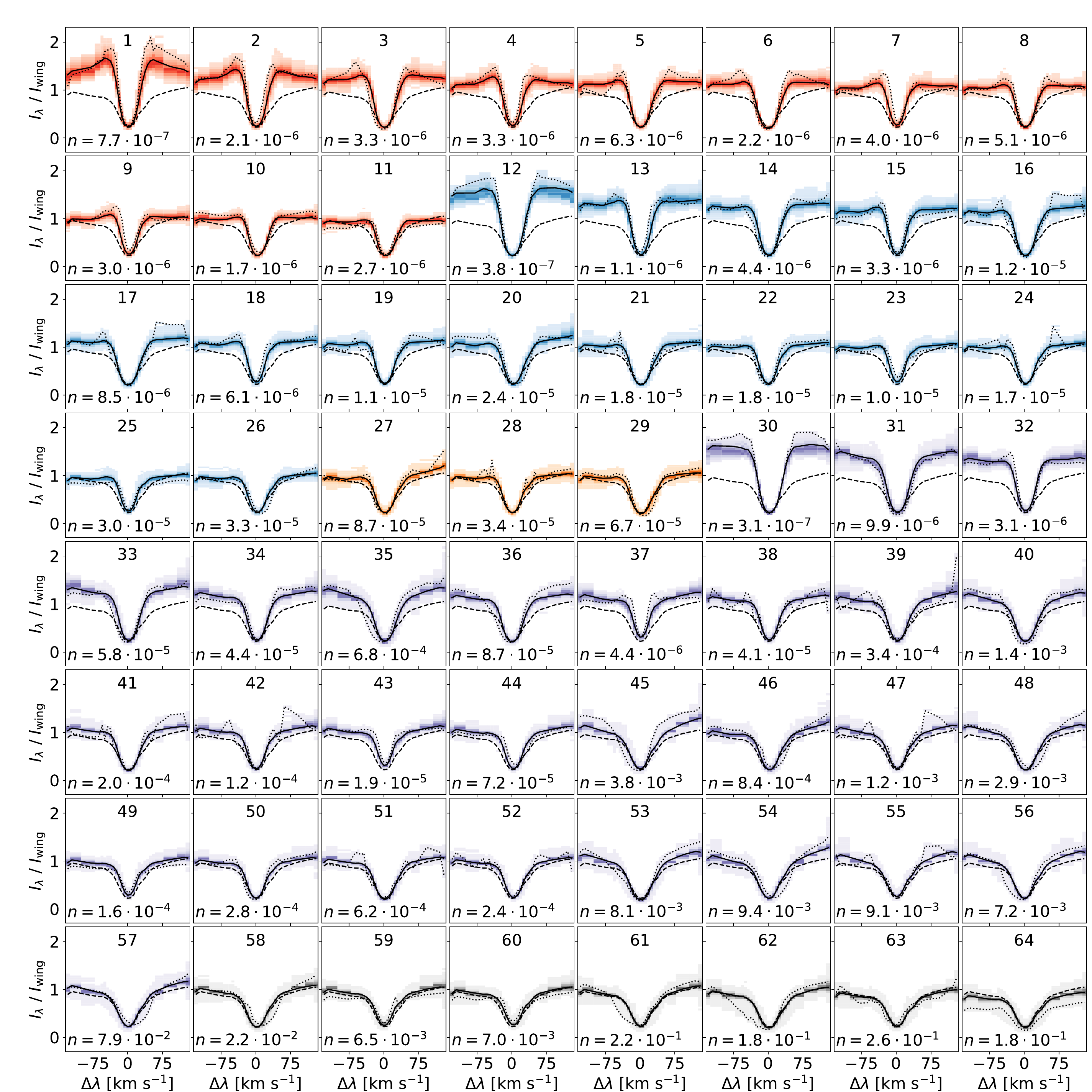}
            \caption{
            Overview of all the clusters from our \(k\)-means model.
            The solid lines represent the cluster centres of the respective clusters.
            The shaded areas represent the spectral line density within each cluster, where a stronger colour means a higher density;
            red and blue represent QSEBs, where the cluster centres in the red group have \(\mean{I}_{\lambda_{\text{iw}}} > \mean{I}_{\lambda_{\text{ow}}}\) in both wings, while the cluster centres in the blue group have \(\mean{I}_{\lambda_{\text{iw}}} > \mean{I}_{\lambda_{\text{ow}}}\) in only the blue wing;
            orange represents clusters containing both QSEB and non-QSEB profiles;
            purple represents clusters with MCs;
            grey represents clusters with neither QSEBs nor MCs;
            the dotted line is the spectral profile that deviates the most from its cluster centre;
            the dashed line is the background profile, which is the average profile over the entire dataset and serves as reference;
            \(n\) represents the fraction of pixels from the entire dataset (\(N = 685 \cdot 10^{6}\) pixels) within the respective cluster centre.
            \(I_{\text{wing}}\) is the average intensity of the second to outermost wavelength points in the blue and red wing (we did not choose the outermost wavelength points to avoid blends).
            }
            \label{fig: kmeans}
        \end{figure*}
    SST's high-quality data depend on good seeing and SST's adaptive optics system \citep{2024A&A...685A..32S}.
    We used the SSTRED pipeline \citep{2015A&A...573A..40D, 2021A&A...653A..68L} to process the raw data into science-ready data cubes.
    As part of the pipeline, we applied image restoration with multi-object multi-frame blind deconvolution \citep[MOMFBD; ][]{2005SoPh..228..191V}, which finally resulted in a near-diffraction limited imaging.
    Once the CHROMIS and CRISP data were processed, the CRISP FOV (\(59 \arcsec \times \, 59 \arcsec\)) was aligned to CHROMIS.
    As the FOV of the cameras is different, there are parts of CHROMIS' FOV that are not covered by CRISP, as is visible in Fig.~\ref{fig: data}.
    The alignment was done by cross-correlating photospheric features as captured by the wideband (WB) channels of \(\HB\) and \ion{Fe}{I}~6173~\AA.
    The lower resolution images of CRISP (pixel scale \(0 \farcs 058\)) were aligned to CHROMIS spatially by linear interpolation and temporally by nearest neighbours.
    The CHROMIS WB channel has a filter centred at 4846~\AA\ with bandpass 6.5~\AA.
    The wavelength offset is sufficiently large from the \(\HB\) line that this can be considered to be a photospheric continuum channel. 
    The CRISP WB 6173~\AA\ channel is centred on 6173.8~\AA\ with bandpass 4.5~\AA\ and also shows a photospheric scene.
    \begin{figure*}
    \centering
      \includegraphics[width=17cm]{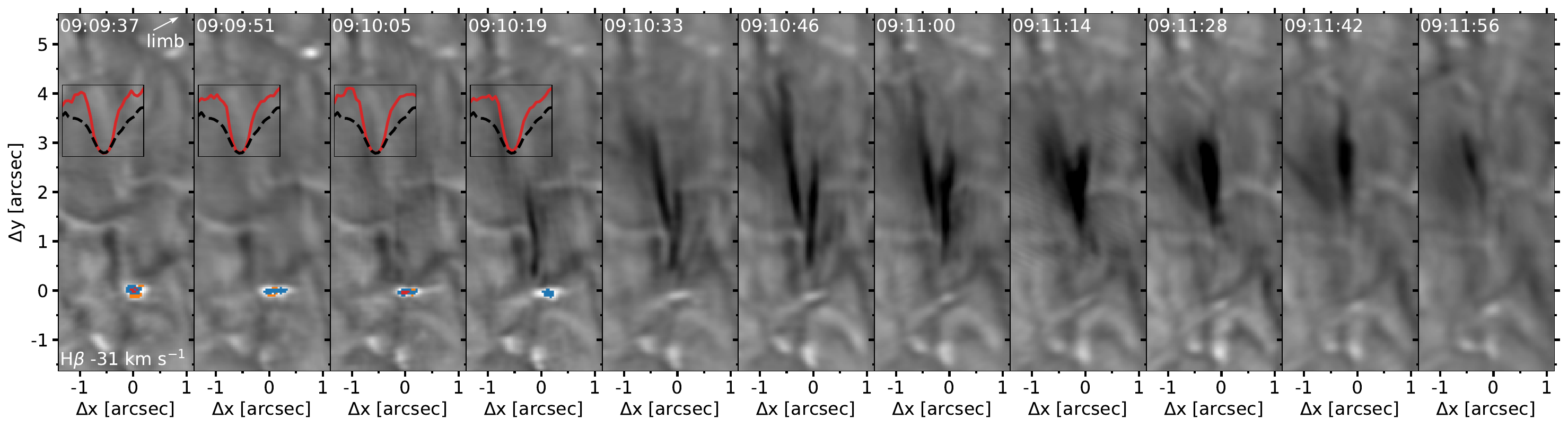}
        \caption{
        Example of spicules connected to a QSEB at their footpoint.
        The panels show a 2-minute sequence of H-beta blue wing images at half the cadence of the original data.
        The arrow in the upper right corner of the first panel shows the direction of the closest limb.
        The coloured pixels mark QSEB detections by the \(k\)-means model, following the colour scheme of Fig.~\ref{fig: kmeans}.
        The small panel in the first four frames show the strongest QSEB profile per frame (red) and the background (black dashed).
        A movie of this event is available in the online material (see \url{http://tsih3.uio.no/lapalma/subl/qseb_spic/sand_qseb_spic_fig03.mp4}).
        }
        \label{fig: burst}
    \end{figure*}
    %
    

\section{QSEB detection and data analysis}
\label{methods}


    \subsection{QSEB detection}

        To detect QSEBs in an automated manner, we implemented SciPy's \citep{2020NatMe..17..261V} \(k\)-means clustering algorithm: a machine learning technique that separates \(m\) data points with \(n\) features into \(k\) clusters by minimising the inertia \citep{1056489, e4fc5e3f-57ed-3636-9789-4190a319a2fa, zbMATH03340881}.
        This technique has been proven useful not only for the detection of QSEBs \citep{2020A&A...641L...5J,2022A&A...664A..72J}, but also for the classification of Stokes profiles 
        \citep{Moe_etal:2023} 
        and spectral profiles in phenomena such as UV bursts \citep{Kleint_Panos:2022}, micro-flares \citep{Testa_etal:2023}, spicules \citep[][]{Bose_etal:2019,Bose_etal:2021a,Bose_etal:2021b}, and surges \citep[][]{Nobrega-Siverio_etal:2021}, among others.
        In our case, the \(m\) data points are the image pixels, while the \(n\) features are the wavelength points along the \(\HB\) spectral line. 
        The number of data points, \(m\), is vast.
        In our datacube covering the \(1818 \times 1194\) pixel FOV and 348 timesteps, there are more than \(685 \cdot 10^{6}\) pixels, even after neglecting the padding around the rotating FOV.
        Hence, we chose a subset of data points for training following \citet{2022A&A...664A..72J}.
        The primary concern when training the algorithm was to separate QSEBs from strong-field magnetic concentrations (MCs) in the photosphere.
        QSEBs differ from MCs by the shape of the wings in \(\HB\), where QSEBs have the moustache-like profile (inner wing emission) that is characteristic for EBs, and MCs have monotonically rising wings that merge with an enhanced continuum-level (see Appendix \ref{app: qseb_vs_mc} for a detailed explanation).
        We defined a line profile to represent a QSEB if the average intensity of the intermediate line wing (\(\mean{I}_{\lambda_{\text{iw}}}\); offset between 43 and 55 \(\kms\)) in either of the wings was larger than the outer line wing (\(\mean{I}_{\lambda_{\text{ow}}}\); outer three wavelength points), that is, where \(\mean{I}_{\lambda_{\text{iw}}} > \mean{I}_{\lambda_{\text{ow}}}\).
        We defined a profile to represent an MC if the profile showed \(\mean{I}_{\lambda_{\text{iw}}} < \mean{I}_{\lambda_{\text{ow}}}\) in both wings.

        We started training \(k\)-means clustering with a dataset that contained pixels \emph{only} with MC and QSEB-like \(\HB\) spectra. 
        In this initial \(k\)-means model, we empirically chose \(k = 28\) (we call this the original model). 
        We also appended the background profile, which is the average profile over the entire dataset, to this model, as the model was based purely on QSEBs and MCs. 
        Even with a highly biased training set, many MC pixels were classified into clusters with QSEB-like cluster centres.
        This is undesirable since we want to identify QSEB pixels uniquely.
        Therefore, we examined each original cluster through a process of sub-clustering to single out which cluster contained a mixture of QSEB and MC profiles.
        We went through each original cluster individually and subclustered them into 64 clusters.
        We retained the original cluster centre if it was clean: if all subcluster centres resembled either QSEB-like or MC-like profiles. 
        For the original clusters where the sub-clustering resulted in a mix of both QSEB and MC-like centres, we created new cluster centres.
        The new cluster centres were created by averaging profiles in all subclusters with similar centres, such as QSEB-like centres with enhanced blue wings, red wings, or both, and MC-like centres.
        After going through the sub-clustering of all the original clusters, we found that 15 original clusters contained both QSEB and MC pixels. 
        From these 15 original clusters, we have generated 44 new cluster centres which were then appended to our original \(k\)-means model. 
        Through the process mentioned above, we arrived at a total of 57 clusters.
        Lastly, we performed \(k\)-means on the background pixels, which are the pixels not classified as QSEBs or MCs.
        For \(k\)-means training on the background, we set \(k = 7\), giving a total of 64 clusters.
        We concluded that this was a manageable number of clusters that allowed us to identify and highlight QSEBs in our dataset effectively. 
        The final \(k\)-means model is presented in Fig.~\ref{fig: kmeans}.
        To highlight QSEBs in the time series, we first made a binary mask marking all pixels in the red and blue clusters (see Fig.~\ref{fig: kmeans}).
        Then, we highlighted all pixels within this binary mask using 26-neighborhood pixel connectivity on the orange clusters.
        A 26-neighbourhood pixel connectivity means that we consider a pixel within the orange group as part of a QSEB if any of its faces, edges, or corners touch a pixel from the red or blue group.
        While this $k$-means identification method is intricate and has some differences compared to the method employed by 
        \cite{2020A&A...641L...5J,
        2022A&A...664A..72J} and
        \cite{2024A&A...689A.156B}, our prime interest was to highlight QSEBs in our data for an effective manual search for connections with spicules.
        After thoroughly inspecting the spectral profiles within the individual clusters used to detect QSEBs, we are confident that the detected pixels within all the clusters in the red and blue groups show clear EB-like emission features in the line wings. 
        This means that we did not find any pixels with an MC signature while inspecting the red and blue clusters.
    \begin{figure*}
    \centering
      \includegraphics[width=17cm]{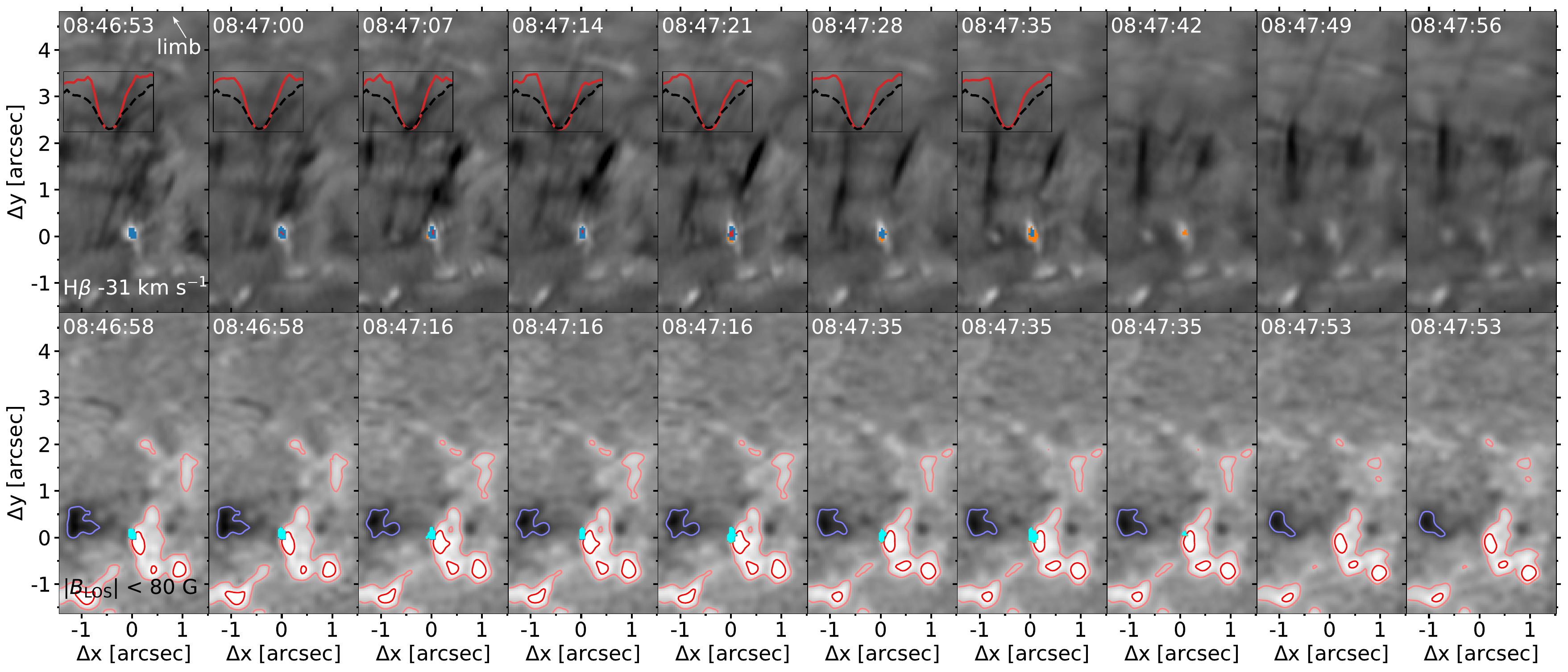}
        \caption{
        A second example of a spicule connected to a QSEB, along with the evolution of the photospheric magnetic field.
        Top row: Sequence of H-beta wing images at 7~s cadence, in the same style as Fig.~\ref{fig: burst}. 
        Coloured pixels mark QSEB detections, following the colour scheme of Fig.~\ref{fig: kmeans}. 
        Bottom row: Evolution of the magnetic field \(\BLOS\).
        The dark red contours encapsulate absolute field strengths stronger than 80~G, and light red and blue contours encapsulate $|B_\mathrm{LOS}| \ge $~40~G ($\sim7\sigma$).
        The cyan colour marks the QSEB pixels detected by the \(k\)-means model.
        Several of these frames are duplicates, as the cadence of the CHROMIS dataset is more than twice as high compared to the CRISP dataset.
        A movie of this event is available in the online material (see \url{http://tsih3.uio.no/lapalma/subl/qseb_spic/sand_qseb_spic_fig04.mp4}).}
        \label{fig: single}
    \end{figure*}
    %

    \begin{figure}
    \resizebox{\hsize}{!}{\includegraphics{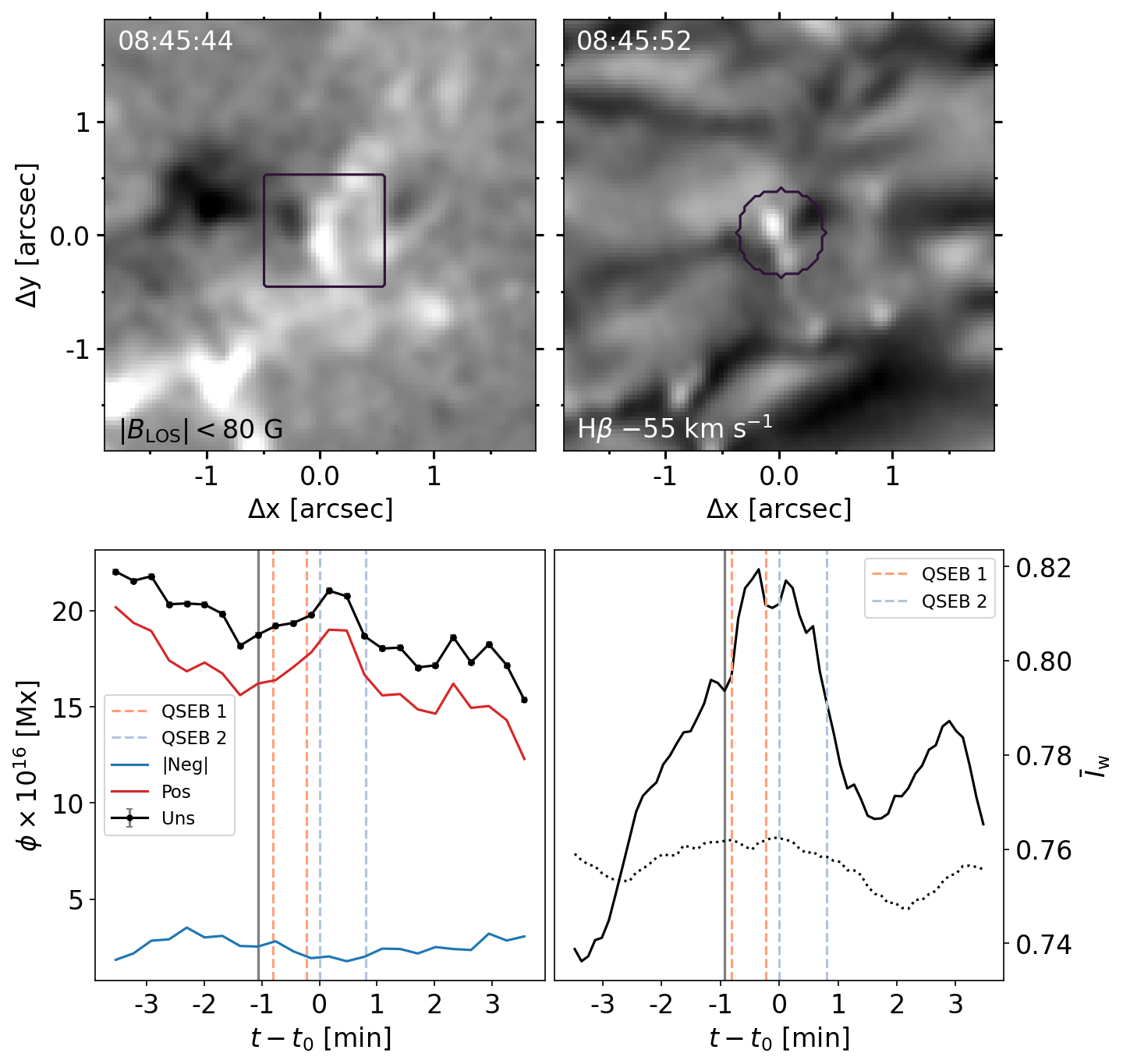}}
    \caption{
        Evolution of the photospheric magnetic field in the area of the QSEB in Fig.~\ref{fig: single}.
        The top left panel shows the magnetic field $B_\mathrm{LOS}$ at a time just before the first detection of two successive QSEBs. 
        The bottom left panel shows the evolution of the magnetic flux $\phi$ inside the purple square in the $B_\mathrm{LOS}$ panel. 
        Vertical dashed lines mark the periods of detections of two QSEBs. 
        QSEB~2 is connected to the spicule formation shown in Fig.~\ref{fig: single}.
        The top right panel shows an \(\HB\) wing image. 
        The bottom right panel shows the light curve of the integrated \(\HB\) inner wing intensity 
        on both sides of the line core (offsets between 31 and 62 \(\kms\)) inside the purple circle in the wing image above, normalised to the outer wing intensity of the quiet Sun reference profile.
        The dotted line is the light curve of the full FOV of the above image. 
        The vertical solid lines in the lower panels mark the time of the images in the upper panels.
        An animation of this figure is available in the online material (see \url{http://tsih3.uio.no/lapalma/subl/qseb_spic/sand_qseb_spic_fig05.mp4}). 
    }
    \label{fig: 4x4}
    \end{figure}

    \subsection{Data analysis}

        After highlighting all QSEB-resembling pixels, we searched for connections between QSEBs and spicules.
        The search was done manually, using the CRisp SPectral EXplorer \citep[CRISPEX;][]{2012ApJ...750...22V}.
        We also used the TrackPy library \citep{2023zndo...7670439A} in Python to label and track events through the time series and search for connections. 
        To get connections that were as clear as possible, we chose the following two criteria.
        The first criterion for positive events was that the spicules aligned with the location of the QSEBs within a close proximity in space, so that it was easy to trace the spicules through the opacity gap \citep{2006A&A...449.1209L, 2012ApJ...749..136L} and down to the QSEB.
        The opacity gap refers to the region in the solar atmosphere around the temperature minimum where the temperature is so low that there is not enough energy to excite electrons to the lower level of the hydrogen Balmer transitions (n=2), and there is effectively low opacity.
        This opacity gap makes it more challenging to rule out chance alignments, as it is more difficult to conclude where the footpoint of the spicule truly is.
        One way to better resolve this is to check if the swaying motion of the spicules is happening about the QSEB, as spicules' footpoints appear to be stationary \citep{2012ApJ...759...18P}.
        We neglected events where QSEBs formed close to a spicule's footpoint if the spicule was not traceable directly to the QSEB.
        The other criterion was that the QSEBs and spicules coincided in time.
        We focused on events where the QSEB formed before or simultaneously as the spicule, while events where the spicules formed before a connected QSEB appeared were rejected.
        We did not use a predefined time window to accept or reject connections between spicules and QSEBs.
        Still, after our search for connections was concluded, we found that no events showed a spicule forming 30~s or later, that is, more than 4 timesteps after a QSEB had ended.
    \begin{figure*}
    \sidecaption
      \includegraphics[width=12cm]{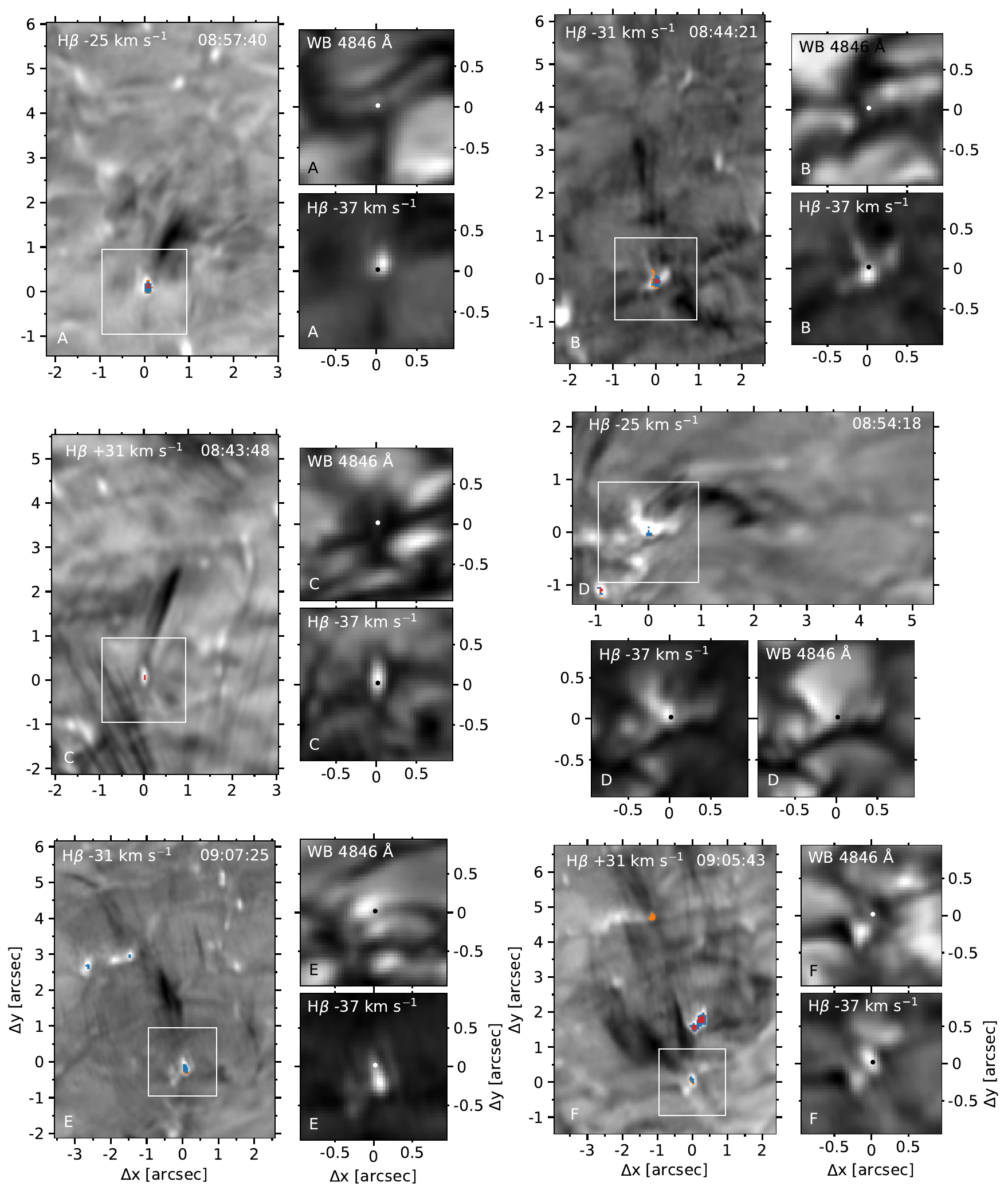}
        \caption{
        Six events, A--F, of QSEBs that are associated with the formation of type~II spicules.
        The big panels show the spicule(s) and their connected QSEB(s).
        Coloured pixels mark QSEB detections.
        Events A, B, D and E are in the \(\HB\) blue wing, while events C and F are in the \(\HB\) red wing.
        The smaller panels show the WB 4846~\AA\ and \(\HB\) blue wing images, zoomed in on the white square in their respective big panels.
        The dots in the small panels are there to highlight the difference in brightening between the WB 4846~\AA\ and \(\HB\) blue wing images.
        All events have accompanying videos in the online material \\
        (event A: \url{http://tsih3.uio.no/lapalma/subl/qseb_spic/sand_qseb_spic_fig06_eventA.mp4}; \\
        event B: \url{http://tsih3.uio.no/lapalma/subl/qseb_spic/sand_qseb_spic_fig06_eventB.mp4}; \\
        event C: \url{http://tsih3.uio.no/lapalma/subl/qseb_spic/sand_qseb_spic_fig06_eventC.mp4}; \\
        event D: \url{http://tsih3.uio.no/lapalma/subl/qseb_spic/sand_qseb_spic_fig06_eventD.mp4}; \\
        event E: \url{http://tsih3.uio.no/lapalma/subl/qseb_spic/sand_qseb_spic_fig06_eventE.mp4}; \\
        event F: \url{http://tsih3.uio.no/lapalma/subl/qseb_spic/sand_qseb_spic_fig06_eventF.mp4}).
        In addition to these six events, the online videos include the events of Figs.~\ref{fig: burst} (see \url{http://tsih3.uio.no/lapalma/subl/qseb_spic/sand_qseb_spic_fig03_3panel.mp4}) and \ref{fig: single} (see \url{http://tsih3.uio.no/lapalma/subl/qseb_spic/sand_qseb_spic_fig04_3panel.mp4}) in this 3-panel format. 
        }
        \label{fig: 6_events}
    \end{figure*}
    %
        

\section{Results}
\label{results}

    The \(k\)-means model gave us a total of 1737 QSEB events, where 80 were visually easy to connect to the formation of one or more spicules.
    For these examples, the QSEBs and spicules coincided well in time, and it was easy for the eye to trace the spicule through the opacity gap down to the QSEB.
    We found that the distance from the top of the QSEB to the bottom of the spicule, ranged between 0.2 to 1.1 Mm, with an average of 0.6 Mm and a standard deviation of 0.2 Mm.
    Apart from these 80 clear cases, we also found over 500 events where the connection was not as clear due to, for instance, high spicule activity or periods with worse seeing.
    Most of these \(> 500\) events look promising, while some of these events showed QSEBs forming at the footpoints of spicule bushes, but a clear, unambiguous connection was impossible to make.
    As a result, we cannot rule out the possibility that many of these 500 cases may have a connection between QSEBs and spicular activity. 
    From the events where we interpreted a clear connection, we noted that QSEBs were connected to the formation of either one, two, or even a burst of several spicules.
    All spicules connected to a QSEB appeared to be fading out of the narrowband \(\HB\) images; this was thoroughly investigated over all wavelength points, as spicules have been shown to disappear prematurely from a fixed bandpass due to changes in the Doppler shift \citep{2016ApJ...824...65P}.
    Out of the most evident examples of QSEBs associated with the formation of spicules, we will present two cases for a more detailed analysis.
    Our first event is presented in Fig.~\ref{fig: burst} (and associated animation).
    This figure shows the formation of a QSEB, followed by two spicules.
    The QSEB forms in the internetwork and shows a clear, flame-like morphology and dynamics, as is evident from the corresponding video.
    The spicules both have a clear ascending phase before they fade out of the images, and we can easily trace them down to the location of the QSEB.
    The first spicule has an apparent speed of 54~\(\kms\) and the second 34~\(\kms\).
    This event occurred outside the CRISP FOV, meaning that we do not have magnetic field information.
    We present our second event in Fig.~\ref{fig: single} (and associated animation), which also shows a QSEB connected to the formation of a spicule.
    Similarly to our first example, we can easily trace the spicule down to the location of the QSEB, and it appears to fade out of the images at the end of its lifetime.
    The \(\BLOS\) frames show a stronger positive polarity field, meeting a weaker negative field at the time and location of the QSEB.
    The evolution of the unsigned magnetic flux, \(\phi\), and the integrated \(\HB\) inner wing intensity around the QSEB event is presented in Fig.~\ref{fig: 4x4} and associated animation.
    Two successive QSEBs were detected in this region, and the spicule illustrated in Fig.~\ref{fig: single} is connected to the second QSEB detected at $t-t_0=0$. 
    The top panels contain the magnetic field and the \(\HB\) blue wing just before the first moment our detection method identified a QSEB. 
    The $B_\mathrm{LOS}$ map in the upper left panel in Fig.~\ref{fig: 4x4} shows two touching opposite polarity patches, and the \(\HB\) wing image shows a clear brightening located at the polarity inversion line. 
    The lower left panel shows that the negative magnetic flux in the area of the purple square decreases during the lifetime of QSEB~1 and the total flux decreases with QSEB~2.
    The QSEB~2 \(\HB\) wing enhancement declines along with the decrease in flux.
    \begin{figure*}
    \centering
      \includegraphics[width=17cm]{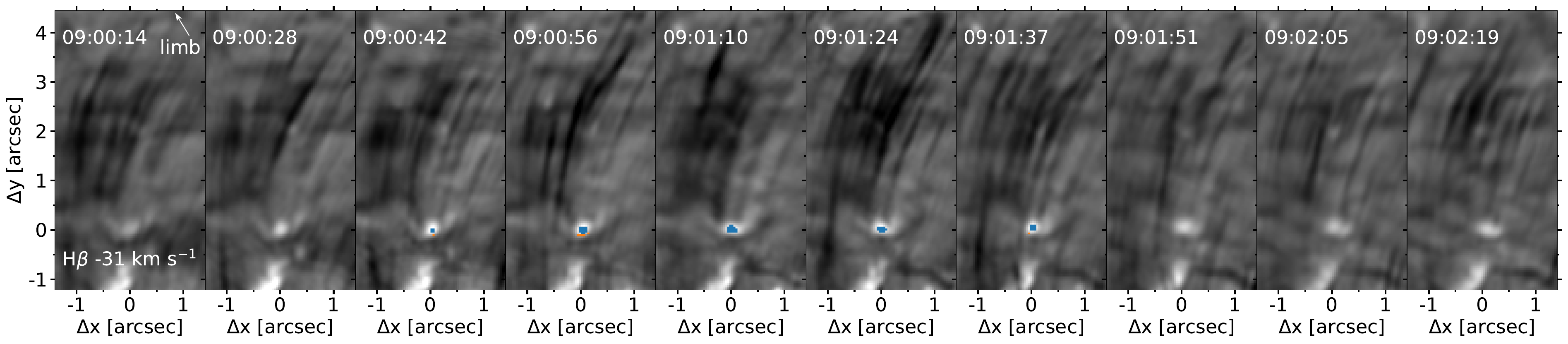}
        \caption{
        Example of high spicule activity connected to a QSEB.
        These panels follow the same convention as Fig.~\ref{fig: burst}.
        An animation of this figure is available in the online material (see \url{http://tsih3.uio.no/lapalma/subl/qseb_spic/sand_qseb_spic_fig07.mp4}). 
        }
        \label{fig: superposition}
    \end{figure*}

    Figure \ref{fig: 6_events} shows snapshots of six other events of QSEBs connected to spicule formation.
    The large panel of event A shows the first of two separate spicules forming after one another, appearing to originate from the same QSEB event.
    Event B shows two parallel spicules forming from a QSEB.
    Event C shows a lonely spicule forming from a QSEB.
    Event D shows many spicules forming from a cluster of QSEBs.
    Event E shows a snapshot of increased spicular activity in the blue wing related to increased QSEB activity.
    Event F shows the magnetic patch just below that of event E, in the red wing.
    For all six events, the smaller panels of the intermediate line wing of \(\HB\) show strong brightening where the QSEB is detected.
    At the same time, this brightening is not visible in the WB 4846~\AA\ images.
    This absence of a bright structure in the WB images validates the QSEB detections.
    All events contain spicules that fade from the \(\HB\) line wing images, meaning that none show a descending phase.
    We could not find a clear descending phase in the other positions of the H-beta line either.
    An example of an event where it is not possible to find a clear, unambiguous connection is presented in Fig.~\ref{fig: superposition}.
    This figure shows high spicule activity coinciding with a QSEB event.
    Many spicules coincide with the QSEB before, during, and after its lifetime.
    There is also a strong MC just below where the QSEB forms.
    It cannot be excluded that this MC is the actual footpoint of the spicules and that the QSEB is unrelated. 
    This presented case is among those \(>500\) events where the connection between a QSEB and spicular activity is ambiguous.
    %


\section{Discussion}
\label{discussion}

    This work aimed to find a connection between QSEBs and spicule formation.
    Finding a clear, unambiguous connection is challenging, mainly because the overcrowded bushes of spicules show complex and fast-changing dynamics.
    Nevertheless, we found 80 clear examples of spicule formation that are connected to QSEBs.
    Most spicules appeared soon after the onset of the QSEBs and none later than 30~s after the ending of the QSEBs. 
    The connected events where the spicules were easy to trace over time showed spicules that faded out of the images over every wavelength point, strongly suggesting that they were all type~II spicules.
    The clearest example of spicule formation connected to a QSEB is presented in Fig.~\ref{fig: burst}.
    The first four frames show the QSEB, as detected by the \(k\)-means clustering.
    Just before the QSEB ends, two spicules rise above it.
    Both spicules appear to rise from the location of the QSEB and fade out of the images over every wavelength point, strongly suggesting that these are type~II spicules associated with the QSEB.
    As QSEBs are considered a tracer for magnetic reconnection \citep{2016A&A...592A.100R, 2018MNRAS.479.3274S, 2020A&A...641L...5J}, we demonstrate a connection of the thermal and kinetic part of magnetic reconnection related to spicules.
    This event is outstanding in our dataset, as it shows clear spicules in the internetwork outside the stronger magnetic field regions of the network.
    There is no superposition with other spicules near this event, making the fading out of the images over every wavelength point straightforward to establish.
    Connecting the spicule through the opacity gap down to the QSEB is also straightforward.
    The QSEB is additionally free of surrounding MCs, making this event an almost perfect example of type~II spicules that appear to be driven by reconnection.
    That is less the case for the second event, where there are several other spicules, as well as another connected QSEB and spicule event occurring just before this respective event.
    However, it is still evident from the upper panels of Fig.~\ref{fig: single} that this event shows robust signs of a type II spicule driven by magnetic reconnection similar to the event of Fig.~\ref{fig: burst}.
    The lower panels show the evolution of the magnetic field and that the QSEB occurs at the interface between a strong positive polarity patch and a weaker negative polarity. 
    The interaction between these opposite polarities at the location of the QSEB is shown in more detail in Fig.~\ref{fig: 4x4} and associated animation, where we also show the evolution of the magnetic flux in the area covering the QSEB.
    The unsigned flux decreases after the onset of the QSEB which is mostly due to the decrease of the stronger positive field at the location of the QSEB (we notice that the decrease in the unsigned flux in the last time frames is due to the positive patch leaving the box and is not related to the QSEB events). 
    Magnetic flux cancellation is also clear from the dynamical evolution of the interacting opposite polarity patches in the animation of the $B_\mathrm{LOS}$ maps. 
    These signs of flux cancellation connected to this particular case are less obvious or even absent in many of our cases.
    Many of the events occur in regions that are dominated by noise in the \(\BLOS\) maps, and similarly to \citet{2020A&A...641L...5J}, we find several events in unipolar magnetic patches.
    Compared to their work, we observe closer to the limb ($\mu=0.57$) so that line-of-sight effects are even more substantial, and magnetic field patches in deep intergranular lanes may be hidden behind the granular ``hills'' in the foreground.
    This projection effect did not play a role in the observations of \citet{2019Sci...366..890S}, as they observed close to the disk centre at \(\mu = 0.995\).
    They reported the appearance or emergence of small opposite polarity flux near a stronger magnetic field in a network patch, which they connected to the observation of enhanced spicule activity. 
    They suggested that this increase in spicule activity resulted from the weak opposite polarity field interacting with the dominant pre-existing field.
    Their high spatial resolution data showed large-scale patterns in the evolution of the photospheric magnetic field and corresponding spicule activity, supporting the reconnection hypothesis.
    However, the spectral resolution of the \(\Ha\) data and the temporal resolution of the magnetic field data make it more difficult to find direct evidence of magnetic reconnection driving spicules on a more detailed and individual basis. 
    Our work provides direct evidence of magnetic reconnection driving a subset of spicules through the detection of QSEBs at their footpoints, strengthening the argument for reconnection as a key process in spicule dynamics.
    Out of the 1737 QSEBs, we found that 80 were clearly connected to spicule formation. 
    In addition, for about 500 QSEBs, the connection was suggestive but inconclusive. 
    This raises the idea that many events may show connections due to chance alignment, which is not possible to rule out in our observations.
    However, the examples we present strongly suggest true connections between QSEBs and spicules linked to the same event of magnetic reconnection.
    The event presented in Fig.~\ref{fig: burst} is located in a particular quiet region with little to no spicular activity.
    Then a clear QSEB appears followed by the formation of two strong spicules, where an aligned connection is clear.
    After the spicules fade out of all the passbands, the region returns to a dynamical activity level similar to that before the QSEB and cotemporal spicules occurred, strongly suggesting that the consecutive dynamics and alignment between the QSEB and spicules are not coincidental.
    Another quite suggestive event is presented in Fig.~\ref{fig: 6_events}, panel D, which shows a cluster of QSEBs leading to a cascade of spicules.
    Panels E and F of the figure show that increased QSEB activity leads to increased spicular activity.
    The other events presented in this work show similar suggestive connected dynamics between QSEBs and spicules.
    That said, we observe that most spicules are not connected to a corresponding QSEB.
    The best-seeing frames in our data show, on average, 323 spicules and 70 QSEBs, which means that for every QSEB visible, there are 4.6 spicules.
    This average number of spicules in our data is based on the counting of dark streaks in blue and red wing \(\HB\) images, which makes the number of spicules a lower limit. 
    Our data shows that most QSEBs are not associated with spicules within our criteria, and most spicules do not show any QSEBs at their footpoints.
    There may be several reasons for this, such as:
    \begin{enumerate}
        \item the height of reconnection
        \item the spatial resolution of the data
        \item reduced seeing and weak spicule signal in the less optically thick chromosphere of the \(\HB\) line compared to \(\Ha\)
        \item high spicule activity
        \item more than one driving mechanism for type~II spicules
    \end{enumerate}
    We will discuss these points in detail below.
    \subsection{Reconnection height.}
        For \(\HB\) to get the characteristic EB signature, which is emission in the intermediate line wings with a dark absorption core, the reconnection and associated plasma heating must be located sufficiently low in the atmosphere.
        If a spicule is driven by reconnection occurring at a height where it is not traceable by \(\HB\), our data shows the spicule but cannot connect it to an associated QSEB.
        Many type II spicules may be driven by reconnection without visible \(\HB\) QSEBs at their footpoints. 
        The idea that \(\HB\) QSEB wing emission only serves as a reconnection proxy for a limited extent of the atmosphere is supported by the comparative study on QSEBs in \(\HB\) and \(\He\) by \citet{2024A&A...683A.190R}.
        They report that while most QSEBs are visible in both lines, a relatively large fraction of QSEBs are only detected in either \(\HB\) or \(\He\).
        The \(\He\) line forms higher than the wings of the \(\HB\) and \(\Ha\) lines
        \citep{2023A&A...677A..52K},
        implying that the QSEB signature in the \(\He\) line core is formed higher in the atmosphere than the \(\HB\) wing QSEB signature.
        These single-line detections indicate that some reconnection events occur deeper in the atmosphere and are only visible in \(\HB\), while some other reconnection events occur higher and are only visible in \(\He\). 
        Possibly, reconnection events that occur just above the \(\HB\) wing formation heights and are visible as the \(\He\) QSEBs can be connected to spicule formation. 
        This is an open question that should be pursued in future studies.
    \subsection{Spatial resolution.}
        QSEBs are such small events that just going from \(\Ha\) to the shorter wavelength of \(\HB\) gives many more detected QSEBs \citep{2020A&A...641L...5J}.
        The prediction of finding even more QSEBs with higher spatial resolution was later reported by \citet{2022A&A...664A..72J}, as they found a sharp cutoff of detected QSEBs at the resolution limit of the SST.
        This conclusion was verified when \citet{2024A&A...683A.190R} used the higher spatial resolution in the shorter wavelength \(\He\) to track QSEBs, and they report that they found 1.7 times more QSEBs in \(\He\) compared to \(\HB\).
        This increased number of detections suggests that we can make more connections between QSEBs and spicules with higher spatial resolution. 
    \subsection{Seeing and weak signal.}
        Both \citet{2022A&A...664A..72J} and \citet{2024A&A...683A.190R} report that the number of detected QSEBs is highly dependent on the seeing, which is a strong correlation that we also observe in this dataset.
        The seeing, and with it the number of detections, varies notably during the observations, and there are several parts where we see a reduction of detections over numerous consecutive frames.
        For example, in one part of the observations, we saw an average of 55 detections over ten consecutive frames before the seeing degraded.
        In the ten subsequent frames, the number of detections was almost halved, with an average of 30 detections.
        In addition to constraining QSEB detection, the seeing also moderately affects the visibility of spicules.
        This is especially relevant for \(\HB\) compared to the more optically thick \(\Ha\) \citep[see, e.g.,][]{2024ApJ...965...15K, 2024UiO...B}, where spicules are thicker, longer, and darker in wing images.
        With the varying seeing combined with the weaker spicule signal in \(\HB\), we find many examples where we may have inferred a QSEB driving a spicule.
        For these cases, the association remains ambiguous since the spicule is too faint. 
        We expect the spicules in these examples would probably have been clearer in \(\Ha\).
        We also found that the method of using \(\HB\) line width maps for spicule detection did not allow for a better tracking of the spicules in our data than \(\HB\) wing maps. 
        \citet{2023ApJ...944..171B} showed \(\HB\) width maps of a region with a coronal bright point but also used wing images for spicule detection.
        For spicule detection in \(\Ha\), on the other hand, width maps are very robust for tracking spicules over time and accounting for changes in effective Doppler shift \citep{2016ApJ...824...65P}.
    \subsection{High spicule activity.}
        Spicules primarily reside in densely crowded bushes associated with magnetic network regions. 
        In these typical spicule environments, tracking spicules over time is challenging due to superposition combined with the effect of spicules swaying in and out of single wavelength passbands.
        If a QSEB occurs in such a network region, the high number of spicules and their complex dynamical evolution often make it nearly impossible to make an unambiguous connection.  
        We present an example of a QSEB forming in a region with high spicule activity in Fig.~\ref{fig: superposition}; this region belongs to the large network patch in the middle of the FOV.
    \subsection{Other driving mechanisms.}
        The abovementioned factors address the limitations that may explain why we find relatively few clear connections between QSEBs and spicules. 
        Another obvious explanation could be that only a fraction of spicules are driven by magnetic reconnection.
        Other mechanisms, such as release of amplified magnetic tension \citep{2017ApJ...847...36M,2017Sci...356.1269M,2020ApJ...889...95M,2017ApJ...849L...7D} or wave-related processes \citep{1999ApJ...514..493K,2004Natur.430..536D,2006ApJ...647L..73H,2011Natur.475..477M}, may drive a significant fraction of spicules. 
        Whether type~II spicules driven by magnetic reconnection or other processes lead to distinctive different dynamical behaviour is still an open question. 
        It is further possible that not all events detected as QSEBs stem from magnetic reconnection. 
        Recently, \citet{2024arXiv241004964C} 
        reported on short-lived events in a network patch that exhibit Ellerman bomb-like enhancements of the \(\Ha\) wings.
        Their spectropolarimetric data acquired with the integral field spectrograph MiHI \citep{2022A&A...668A.149V} 
        showed that the network patch was uni-polar without any clear indications of opposite polarity fields nearby. 
        The authors proposed that the \(\Ha\) wing enhancements in these so-called photospheric hot spots could be caused by convection-driven magnetic field intensification.
        While we cannot rule out this scenario for the \(\HB\) wing enhancement in QSEBs in our data, we observe a distinct flame-like morphology and rapid intensity variability for many QSEBs. 
        The observed area in our data is much closer to the limb than the observations of 
        \citet{2024arXiv241004964C}. 
        This gives a side view that provides a clear detection of elongated QSEB flames.
        Rapid variability and flame morphology are best explained by magnetic reconnection 
        \citep{2011ApJ...736...71W, 
        2013JPhCS.440a2007R, 
        2017ApJ...839...22H, 
        2017A&A...601A.122D}. 

        Our working hypothesis was simple: if a spicule is driven by magnetic reconnection in the photosphere, we would first see a QSEB forming and then a spicule forming from this QSEB.
        We have indeed observed many spicules forming from a QSEB.
        We have also seen many spicules forming simultaneously with a QSEB.
        We also observed some spicules that formed after a very short delay after the corresponding QSEB was no longer detectable.
        All these three categories of QSEB-to-spicule events have been included in our statistics.
        We have also observed some spicules forming before a connecting QSEB was visible.
        We did not consider QSEBs forming after a respective spicule as tracers for reconnection-driven spicules, as they did not align with our simple hypothesis.
        However, the recent work of \citet{2024A&A...683A.190R} suggests that the connection between QSEBs and spicules is more complex than our base hypothesis.
        They report that about one-fifth of QSEBs detected by both \(\HB\) and \(\He\) first occur in \(\He\) before they are visible in \(\HB\).
        They further conclude that these events' reconnection starts higher in the atmosphere and then propagates downwards.
        This conclusion is similar to that of \citet{2024A&A...689A.156B}, who find UV brightenings occurring before a corresponding QSEB for a third of their QSEB-UV brightening events.
        Therefore, there may be events of QSEBs associated with spicule formation where the reconnection site is traceable in \(\He\) before being traceable in \(\HB\).
    \subsection{Future work and conclusions}
        The CHROMIS \(\HB\) observations provide high numbers of QSEB detections which serve as an effective proxy for tracing magnetic reconnection in the deep solar atmosphere. 
        However, \(\He\) detects more QSEBs and may provide valuable insight related to QSEBs connected to spicule formation.
        We also conclude that the \(\Ha\) spectral line is more effective for tracing spicules compared to \(\HB\), as they appear thicker, longer, and with higher contrast.
        To better probe magnetic reconnection over a larger atmospheric range, an observational programme that includes simultaneous measurements of multiple Balmer lines would be most effective for establishing more unambiguous connections between QSEBs and spicules.  
        With the current set of prefilters of the CRISP and CHROMIS instruments at SST, such a programme would include \(\Ha\), \(\HB\), and \(\He\). 
        High spatial resolution is also crucial, and it will therefore be interesting to see what the new generation of ground-based telescopes, like, for instance, the Daniel K. Inouye Solar Telescope \citep[DKIST;][]{2020SoPh..295..172R} and the upcoming European Solar Telescope \citep[EST;][]{2022A&A...666A..21Q}, can find on magnetic reconnection driving spicules.
        In our work on connecting the thermal and kinetic components of magnetic reconnection, we have found several very suggestive events where type~II spicules are connected to QSEBs.
        Even though the small number of positive events may result from the above reasons, we do not claim that magnetic reconnection is the sole driver for type~II spicules.
        The reconnection hypothesis may very well be relevant for only a subset of type~II spicules, as this study's results do not exclude other mechanisms for driving type~II spicules.
        That said, we provide individual connections between type~II spicules and magnetic reconnection.

\begin{acknowledgements}
 
The Swedish 1-m Solar Telescope (SST) is operated on the island of La Palma
by the Institute for Solar Physics of Stockholm University in the
Spanish Observatorio del Roque de los Muchachos of the Instituto de
Astrof{\'\i}sica de Canarias.
The SST is co-funded by the Swedish Research Council as a national research infrastructure (registration number 4.3-2021-00169).
This research is supported by the Research Council of Norway, project numbers 250810, 
325491, 
and through its Centres of Excellence scheme, project number 262622.
J.J. is grateful for travel support under the International Rosseland Visitor Programme. 
J.J. acknowledges funding support from the SERB-CRG grant (CRG/2023/007464) provided by the Anusandhan National Research Foundation, India.
D.N.S acknowledges support by the European Research Council through the
Synergy Grant number 810218 (``The Whole Sun'', ERC-2018-SyG).
S.B. gratefully acknowledges support from NASA's IRIS (NNG09FA40C) and MUSE (80GSFC21C0011) contracts to LMSAL.
A.B.G.M. acknowledges support by the Swedish National Space Agency through the Call 2023-C, project number 31004980.
We made much use of NASA's Astrophysics Data System Bibliographic Services.
\end{acknowledgements}

\bibliographystyle{aa} 
\bibliography{bibtex/sources,bibtex/ref_daniel}

\begin{appendix}
    \section{Differences between QSEBs and MCs}
    \label{app: qseb_vs_mc}
        \begin{figure*}
        \centering
          \includegraphics[width=17cm]{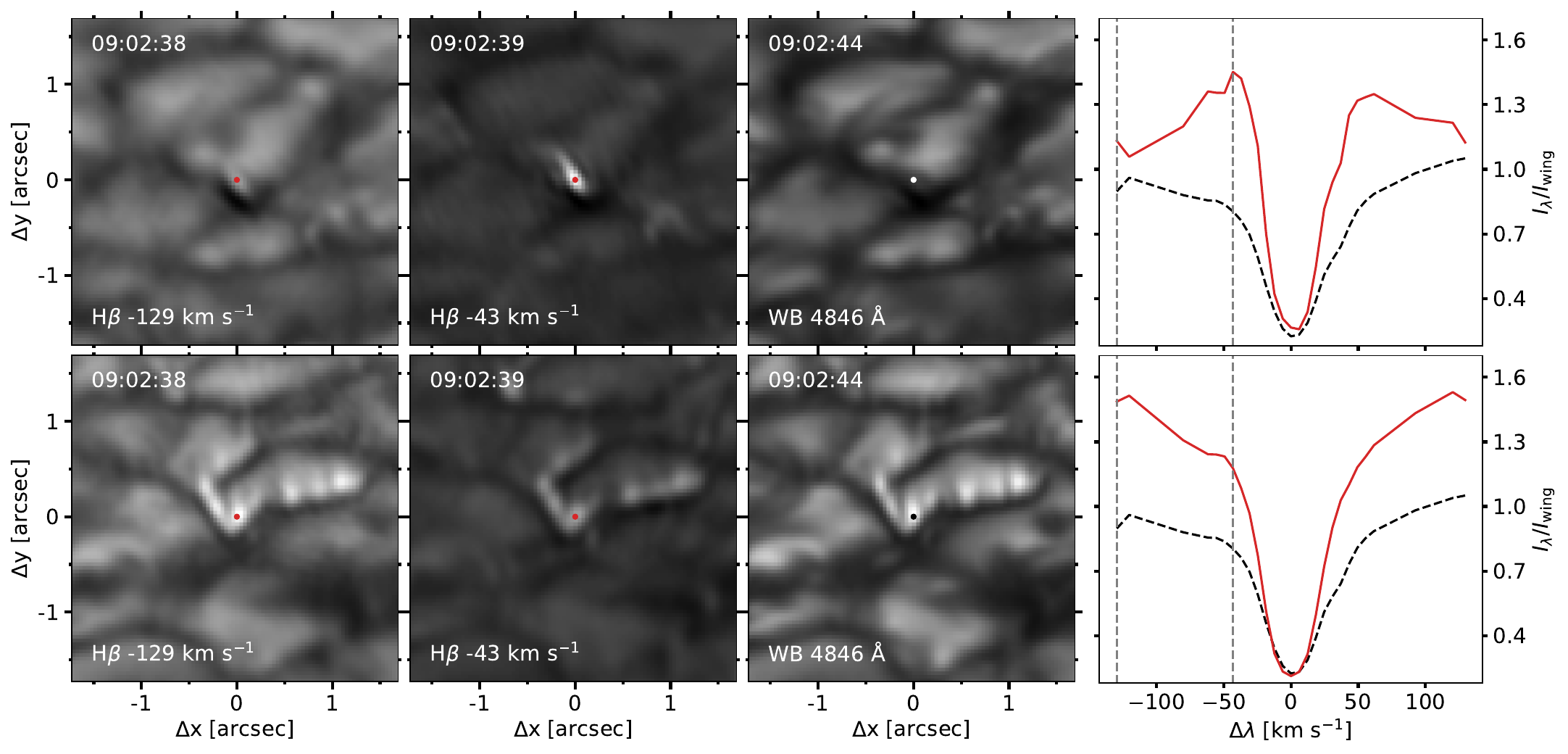}
            \caption{
            Two separate regions of the FOV showing a QSEB in the upper panels vs a cluster of MCs in the lower panels.
            The first two columns show the region of the QSEB and MCs as shown by the narrowband filter images; the labels in the lower left corners define the wavelength positions of the images.
            The third column shows the regions of the QSEB and MCs as shown by the wideband filter images.
            Each column follows the same scaling.
            The last column shows the spectral profiles for the separate features (red solid) and the background profile (black dashed); the vertical lines represent the wavelength positions for the two first columns.
            The spectra are acquired from the pixel marked by the red dot in the centre of the images.
            The white and black dots in the wideband images help guide the eye.
            A movie of these features is available in the online material (see \url{http://tsih3.uio.no/lapalma/subl/qseb_spic/sand_qseb_spic_figA1.mp4}).
            }
            \label{fig: qseb_vs_mc}
        \end{figure*}
        To illustrate the distinctive differences between QSEBs and MCs, Fig.~\ref{fig: qseb_vs_mc} shows details of a QSEB and a cluster of MCs; the QSEB and MCs are taken from the same time frame but separate regions within the FOV.
        The upper panels show that the QSEB is visible in the line wing of \(\HB\) but not at the outermost wavelength point nor in the WB 4846~\AA\ image.
        The MC cluster is visible in both the narrowband and the WB 4846~\AA\ images.
        The QSEB has an elongated morphology with flame-like dynamics, as the corresponding movie shows, while the MCs do not show this dynamic behaviour.
        In addition to their distinct visibility and extended, flickering dynamics, QSEBs are different from MCs by the shape of their spectral profiles, as shown by the last column.
        The upper panel shows the typical EB signature: emission features in the line wings and a nearly unaffected line core, traditionally called a moustache-like profile.
        The lower panel shows the typical MC signature, that is, enhanced line-wing intensity, monotonically increasing from the line core to the continuum (the decrease in intensity seen in the outermost wavelength points is due to spectral blends).
        However, separating one from the other is not always as obvious as in this presented case.
        There is a grey area between QSEB and MC spectral definitions.
        We often observe small bright features that show flickering, QSEB-like behaviour, but the line wings do not have the typical QSEB emission features.
        In these events, the spectral line wings are closer to flat (see, for example, cluster centres 30 and 32 in \ref{fig: kmeans}); they neither show line wing emission nor monotonically rise towards the continuum.
        As these features show QSEB-like behaviour but lack the spectral signature, one may argue that these are likely QSEB events where the emission did not get strong enough to surpass the intensity in the outer-most line positions.
        However, we were only interested in the clearest indications of QSEBs connected to spicule formation: detections with flame-like morphology and clear emission features in the line wings.
        Therefore, we neglected any event that could be categorised in the grey area between the spectral signatures for QSEBs and MCs.

\end{appendix}

\end{document}